# Evidence for a pressure-induced antiferromagnetic quantum critical point in intermediate valence UTe$_2$

**Authors**
S. M. Thomas,[1*] F. B. Santos,[1,2] M. H. Christensen,[3] T. Asaba,[1] F. Ronning,[1] J. D. Thompson,[1] E. D. Bauer,[1] R. M. Fernandes,[3] G. Fabbris,[4] P. F. S. Rosa[1]

**Affiliations**
1. Los Alamos National Laboratory, Los Alamos, NM, USA
2. Escola de Engenharia de Lorena, Universidade de Sao Paulo (EEL-USP), Materials Engineering Department (Demar), Lorena – SP, Brazil
3. School of Physics and Astronomy, University of Minnesota, Minneapolis, MN 55455, USA
4. Advanced Photon Source, Argonne National Laboratory, Argonne, IL 60439, USA

*. Corresponding author, smthomas@lanl.gov

**Abstract**

UTe$_2$ is a recently discovered unconventional superconductor that has attracted much interest due to its potentially spin-triplet topological superconductivity. Our ac calorimetry, electrical resistivity, and x-ray absorption study of UTe$_2$ under applied pressure reveals key new insights on the superconducting and magnetic states surrounding pressure-induced quantum criticality at P$_{c1}$ = 1.3 GPa. First, our specific heat data at low pressures, combined with a phenomenological model, show that pressure alters the balance between two closely competing superconducting orders. Second, near 1.5 GPa we detect two bulk transitions that trigger changes in the resistivity which are consistent with antiferromagnetic order, rather than ferromagnetism. Third, the emergence of magnetism is accompanied by an increase in valence towards a U$^{4+}$ (5f$^2$) state, which indicates that UTe$_2$ exhibits *intermediate valence* at ambient pressure. Our results suggest that antiferromagnetic fluctuations may play a more significant role on the superconducting state of UTe$_2$ than previously thought.

## Introduction

Spin-triplet superconductors have recently attracted renewed interest due to their potential topological properties (*1*). UTe$_2$ is a newly discovered superconductor that has been argued to host spin-triplet pairing due to a large H$_{c2}$ that violates the paramagnetic limit (*2*, *3*). Nuclear magnetic resonance measurements revealed a very small change in the Knight shift in the superconducting state, which is consistent with the spin-triplet scenario (*4*). Under magnetic field applied along the *b* axis, re-entrant superconductivity was discovered, which abruptly changes to the normal state at a metamagnetic transition at 34.5 T (*5*). In addition, scanning tunneling microscopy experiments found evidence of in-gap states argued to be evidence of chiral superconductivity (*6*). Finally, Kerr effect measurements revealed field-trainable time-reversal symmetry breaking in the superconducting state, which is consistent with topological (Weyl) superconductivity (*7*).

Applied pressure is a clean, symmetry-preserving tuning parameter that may shed light on the spin-triplet superconducting state of UTe$_2$. Prior hydrostatic pressure work found evidence for two superconducting transitions above 0.3 GPa (*8*). How these transitions extrapolate to zero pressure remains an open question. As pressure is increased, putative magnetic order emerges and superconductivity is rapidly suppressed. Notably, no superconducting transition was found to occur in the magnetically ordered state. Conversely, another pressure study argued that, at the pressure where magnetic order emerges, there is heterogeneous coexistence of magnetic and superconducting states (*9*). In this scenario, the superconducting regions do not percolate at zero field, and the low-temperature resistance is finite. Application of magnetic field suppresses the magnetic order and enhances superconductivity, which causes a zero-resistance state to reemerge under magnetic field. The discrepancy between these two results in the high-pressure region invites a close evaluation of the phase diagram. More recent pressure studies added two new pieces of information. First, above 0.5 GPa, there is field-reinforced superconductivity for magnetic fields applied along the *a* axis (*10*, *11*). Second, there may be a link between the ambient pressure field-induced metamagnetic transition at 34.5 T and the magnetic state induced with pressure (*12*).

Here we perform electrical transport, ac calorimetry, and x-ray absorption measurements in $UTe_2$ under hydrostatic pressure. We find that the superconducting transition temperature is maximized near a putative antiferromagnetic quantum critical point occurring at a pressure of $P_{c1} = 1.3$ GPa. Similar to prior works, we clearly observe two superconducting transitions that have an opposite pressure dependence. Our results, however, reveal a missing piece in the puzzle: the onset temperatures of the two superconducting states cross at very low pressures. Our phenomenological model shows that these closely lying order parameters compete at atmospheric pressure. Applied pressure favors one superconducting state over the other, but it likely preserves a low-temperature phase in which both orders coexist microscopically and break time-reversal symmetry in the low-pressure regime.

Notably, we also find clear thermodynamic evidence for two phase transitions consistent with *antiferromagnetic* order: $T_{m1}$ sets in near 1.45 GPa, whereas $T_{m2}$ sets in at a slightly higher pressure of 1.51 GPa. The electrical resistivity displays a clear upturn at $T_{m1}$, which is usually a signature of antiferromagnetic order rather than ferromagnetic order (*13*, *14*). In fact, this region of the phase diagram most closely mirrors that of antiferromagnetic $CeRhIn_5$ (*15*), instead of known ferromagnetic superconductors (*16*). Further, the emergence of magnetism is accompanied by an increase in valence towards a $U^{4+}$ ($5f^2$) state, which indicates that $UTe_2$ exhibits intermediate valence at ambient pressure. The increase in valence is accompanied by a decrease in the Kondo coherence temperature, which is generally expected to drive the system from superconducting to magnetic (*17*). Our results provide evidence that one of the two nearly-degenerate superconducting instabilities is strongly enhanced upon approaching an antiferromagnetic transition, which raises important questions about the proposed spin-triplet nature of the pairing state in $UTe_2$.

**Results**
Figure 1 summarizes our ac calorimetry and electrical resistivity data collected at representative pressures. Heat capacity at ambient pressure, shown in Fig. 1(A), displays a peak near $T_{c2} = 1.65$ K, which is consistent with the offset from a zero-resistance state. This main peak is followed by a shoulder in heat capacity occurring at slightly lower temperature, near $T_{c1} = 1.45$ K. Though evidence for two transitions at zero pressure has not been uniformly reported in all prior publications, it has been recently observed in detailed heat capacity measurements (*7*). The presence of two peaks and time-reversal symmetry breaking was taken as evidence for a non-unitary two-component order parameter because of the orthorhombic crystal symmetry of UTe$_2$. As shown in Ref. (*7*), the splitting between these two transitions is small and sample dependent, which explains why this feature may have been missed in earlier reports.

At low pressures, these bulk transitions have opposite pressure dependence and cross at 0.2 GPa. As shown in Fig. 1(B), the small shoulder at T$_{c1}$ at zero pressure moves to higher temperature as pressure is increased and also gains more entropy relative to the transition at T$_{c2}$, which is suppressed fairly linearly with pressure and loses entropy. Further, the zero-resistance state always occurs at the higher of these two transition temperatures. Because the transitions cross, our result indicates that both arise from superconductivity, in agreement with Ref. (*8*).

These observations are consistent with a scenario in which pressure tunes the balance between two closely competing superconducting instabilities. To expand this analysis, we consider the Landau free-energy expansion of two superconducting order parameters that transform as two different one-dimensional irreducible representations of the orthorhombic group, $\psi_1$ and $\psi_2$:

$$\mathcal{F} = \alpha(T-T_c)\psi_1^2 + \alpha(T-T_c)\psi_2^2 + \beta_1\psi_1^4 + \beta_2\psi_2^4 - 2g\psi_1^2\psi_2^2 + \epsilon\left(\psi_2^2 - \psi_1^2\right)$$

Here, $\epsilon$ represents hydrostatic pressure. Experimentally, $\epsilon = 0$ therefore corresponds to P = 0.2 GPa, in which case the two superconducting transitions are accidentally degenerate. Note that $\epsilon > 0$ favors $\psi_1$, whereas $\epsilon < 0$ favors $\psi_2$. Because previous experiments reported time-reversal symmetry breaking at ambient pressure (*7*), the Landau parameters are constrained to $g_1 > 0$ and $(g_1 - g_2/2)^2 < \beta_1\beta_2$. As shown in the supplementary material, if the only effect of pressure is to change the transition temperatures of the two superconducting states, the sum of the specific heat jumps divided by each T$_c$, $\Delta$C$_1$/T$_{c1}$ + $\Delta$C$_2$/T$_{c2}$ (as well as each term in the sum individually) should be constant under pressure. This would not be the case if pressure were to significantly change the quartic coefficients of the Landau expansion, which could in turn affect the time-reversal symmetry breaking nature of the low-temperature state. We note that our ac calorimetry measurements do not provide quantitative values. Nonetheless, the jumps in specific heat at a given temperature can be compared as a function of pressure as we do not expect the extrinsic contribution from the pressure medium to change drastically in the pressure and temperature range being investigated. As shown in Fig. 1(D), the sum of the specific heat jumps divided by T$_c$ are nearly constant at low pressures (P < 0.7 GPa), consistent with the expectation of the model. This indicates that pressure is mainly affecting the transitions temperatures (*i.e.*, the quadratic terms in the Landau free-energy), suggesting that time-reversal symmetry breaking is likely to take place below the second transition temperature across this pressure range (see Supplementary Section S1 for more details of the calculation). As pressure is further increased, however, the sum of the specific heat jumps divided by T$_c$ also increases. One possible reason for this behavior is the proximity to a magnetic boundary, which is predicted to promote an increase in the specific heat jump in *f*-electron superconductors (*18*). This is the first hint of the proximity of UTe$_2$ to a pressure-induced magnetic state.

At higher pressures, UTe$_2$ does, in fact, develop additional ordered states that likely stem from magnetism. Fig. 1(C) shows that at 1.45 GPa, a new peak appears in heat capacity at

$T_{m1} = 3.8$ K. This temperature has been previously associated with the onset of magnetic order. Notably, the signature of bulk superconductivity is still clearly observed in heat capacity at $T_{c1} = 1.8$ K. The observation of bulk superconductivity occurring below magnetic order is in contrast to prior reports (*8–12*). Compared to 1.40 GPa, however, the superconducting peak is broadened significantly and has less entropy. Similarly, the resistance drop to zero becomes broader at this pressure, although zero-resistance is still reached at a similar temperature to the feature in heat capacity. This bulk-like coexistence extends over only a minute pressure range of less than 0.04 GPa, which is probably why it has not been observed previously. As pressure is further increased, the difference in temperature between the superconducting transition obtained from calorimetry and resistivity begins to grow. This separation is evident in Fig. 1(E) in which there is more than a 0.5 K difference between the two transition temperatures at 1.47 GPa. This indicates that the coexistence of superconductivity and magnetism is likely at the macroscopic scale, with the superconducting volume fraction decreasing below the percolation threshold as pressure moves the system deeper inside the magnetically ordered phase.

At a pressure of 1.51 GPa, a clear second magnetic transition emerges at $T_{m2} = 3.0$ K. The higher temperature magnetic transition temperature increases rapidly at a rate of $dT_{m1}/dP = 16$ K/GPa, whereas the lower magnetic transition temperature shows minimal pressure dependence. At this pressure and beyond, no evidence for a bulk superconducting transition is found via ac calorimetry at any temperature above 70 mK. Remarkably, the resistivity still reaches zero resistance at 1.9 K, as shown in Fig. 1(F). As pressure is increased further, the zero-resistance state is suppressed to zero temperature continuously, also in contrast with prior reports (*9*).

To investigate the relationship between superconductivity and magnetism, we measured the critical current necessary to induce a finite resistance. We highlight the behavior at a pressure of 1.57 GPa, where $T_{c1} = 0.9$ K as determined from critical current measurements. Figure 2(A) shows that at this pressure, the current density ($J_c$) below which there is a zero-resistance state is extremely low, reaching a maximum of 21 mA/cm$^2$ at 0.1 K. The critical current density increases fairly linearly with decreasing temperature below about 0.6 K. Above 0.6 K, $J_c$ saturates to a nearly constant value of just below 1.4 mA/cm$^2$ until no detectable evidence of a zero-resistance state occurs near 0.9 K. The reason for this plateau is not understood. To highlight the role of the current density, the inset of Fig. 2(A) shows resistivity versus temperature plotted at different current densities. The difference between the curves sets in at temperatures as high as 1.5 K. Importantly, sample heating cannot explain the effect. The main source of sample heating is contact resistance to the sample, which is of the order of one Ohm. Even at the highest current density, this results in heating of the order of only 100 pW, which is negligible at these temperatures.

Our results not only provide an explanation for why prior reports did not observe a zero-resistance state in this pressure range (*i.e.*, the measurement current was too high), but also enable us to further investigate the claim of a reentrant superconducting state with applied magnetic field in this pressure region (*9*). To this end, we performed field-dependent measurements at 1.57 GPa using a vector magnet and a large current density (J = 160 mA/cm$^2$). Fig. 2(B) shows the results of these normal-state measurements. Even at low fields (H=0.85 T), the resistivity increases by about 30% for fields applied along the hard [010] axis as compared to fields in the (101) plane, which hints at the tendency to move away from a zero-resistance state for fields applied along the hard axis. At a lower current density ($J = 16$ mA/cm$^2$), the resistivity of UTe$_2$ becomes zero within experimental resolution at 0.85 T for fields in the (101) plane, whereas it is finite when the same field magnitude is applied along the $b$ axis. This field-angle dependence is consistent with a recent report in which no reentrant superconductivity is found for fields applied along the b axis (*12*).

Although superconductivity is enhanced for fields in the (101) plane, we reiterate that at lower current densities a zero-resistance state is obtained for all field directions. At zero pressure, $J_c$ as determined from susceptibility measurements is near 10 kA/cm$^2$ (*19*), a factor of nearly 500,000 times larger than the value of 21 mA/cm$^2$ obtained here. Such a low critical current is inconsistent with bulk superconductivity, which is also supported by the lack of any feature in ac calorimetry. Instead, our results are fully consistent with filamentary superconductivity (*20*). One possibility is that superconductivity is percolating either on the surface or between magnetic domains, which has been observed previously when antiferromagnetic order and superconductivity coexist in the prototypical heavy-fermion superconductor CeRhIn$_5$ (*21*).

Remarkably, the magnetically ordered states occurring above $P_{c2} = 1.4$ GPa seem inconsistent with a simple ferromagnetic phase. Fig. 2(C) shows resistivity versus temperature for several pressures near the emergence of magnetic order. Initially, a slight upward inflection is shown at $T_{m1}$ at 1.45 GPa. As pressure is increased, two magnetic transitions become clear in resistivity, $T_{m1}$ and $T_{m2}$, in agreement with heat capacity measurements. At 1.57 GPa, both $T_{m1}$ and $T_{m2}$ show an upward inflection in resistivity as the sample is cooled. At higher pressures, $T_{m2}$ continues to show an upward inflection, but $T_{m1}$ shows a broad downward feature. As will be discussed later, this temperature dependence suggests antiferromagnetic ordering.

Figure 3(A) shows the pressure-temperature phase diagram constructed from data shown in Figures 1–2. The pressure at which the superconducting transition temperature is maximum coincides with the pressure at which the magnetic order extrapolates to zero temperature, which suggests a putative quantum critical point at $P_{c1} = 1.3$ GPa. Below 1.4 GPa, however, there is no evidence for magnetic order within the superconducting state. This once again is quite similar to antiferromagnetic CeRhIn$_5$, in which magnetic order does not occur at temperatures below the superconducting transition in zero applied field (*15*). Figure 3(B) shows the exponent extracted from taking the logarithmic derivative of $\rho(T) = \rho_0 + AT^n$ after subtracting off $\rho_0$. The residual resistivity term was determined by performing a power-law fit over a small temperature range just above $T_c$ or $T_m$. The resulting plot shows a region of linear-in-temperature resistivity centered on the critical pressure of 1.3 GPa. As shown in Fig. 2(C), at 1.32 GPa the resistivity is linear from just above $T_c$ up to 8 K. As temperature is increased, the resistivity becomes sub-linear due to the Kondo coherence temperature being reduced to 40 K at these pressures (see Fig. S2).

We now turn to the uranium valence in UTe$_2$ under pressure. Figure 4(A) displays the uranium L$_3$ x-ray absorption near edge spectroscopy (XANES) spectra of UTe$_2$ at low pressure as well as UF$_4$ (U$^{4+}$ reference) and UCd$_{11}$ (U$^{3+}$ reference) at zero pressure. Contrary to *4f* systems, it is often difficult to determine the absolute valence of uranium from its L$_3$ edge (*22*). Nevertheless, the peak of the white line of UTe$_2$ is substantially shifted to higher energies compared to $UCd_{11}$ and slightly lower energies compared to $UF_4$, which points to an intermediate valence state at ambient pressure closer to 4+ (*22*). Figure 4(B) compares the uranium $L_3$ XANES spectra of $UTe_2$ at two representative pressures. At 2.5 GPa, a small positive shift in the resonance energy is observed as compared to the lowest pressure point at 0.3 GPa. The uranium absorption edge was modeled using an arctangent step function combined with a Gaussian peak (see Fig. S3). The resulting pressure dependence of the Gaussian peak position can be expressed as a change in uranium valence by using the available 3+ and 4+ references.

A small positive shift in the resonance energy is detected starting above 1.25 GPa, which is in the same pressure range as the onset of magnetic order, as shown in Fig. 4(C). The fact that the shift is positive further suggests that the valence is not an integer (*i.e.*, not fully 4+) at ambient pressure. By taking $UCd_{11}$ and $UF_4$ as $U^{3+}$ and $U^{4+}$ references, respectively, the energy shift at 2.5 GPa implies an apparent reduction of 0.10(4) electrons towards $5f^{\,2}$ ($U^{4+}$). We note, however, that this number is an upper limit because either a shift in the *6d* orbitals without *5f* participation or structural changes as a function of pressure may partially explain the observed behavior.

**Discussion**

Above $P_{c2}$ = 1.4 GPa, two magnetic transitions emerge, the signatures of superconductivity in heat capacity and electrical transport occur at different temperatures, and the uranium valence starts to increase. These experimental observations warrant further discussion. First, it is hard to reconcile two magnetic transitions with ferromagnetic order at zero field, which suggests that the pressure-induced magnetic order is antiferromagnetic. Further evidence for antiferromagnetic ordering can be obtained by considering the temperature dependence of the resistivity at each of the magnetic transitions. For a second-order magnetic transition in a metal, the resistivity is expected to follow the Fisher-Langer behavior, meaning that dρ/dT is proportional to the heat capacity with a positive constant of proportionality (*13*). Thus, the resistivity should decrease as the temperature is lowered through the transition. At an antiferromagnetic transition, however, the resistivity may either show an upward or downward inflection depending on the ordering wave vector Q (*14, 23*). Therefore, there are two possible scenarios to interpret our transport results. The first is that the transitions are ferromagnetic, and the Fisher-Langer scaling behavior is violated. The second, which seems more natural, is that the pressure independent upward inflection at $T_{m2}$ implies antiferromagnetic order with constant Q. As shown in the inset of Fig. 2(C), $T_{m1}$ also shows an upward inflection (i.e., a dip in dρ/dT) at a pressure of 1.57 GPa. When the pressure is further increased, this changes to a downward inflection, as seen at 1.66 GPa. This change in character is also consistent with antiferromagnetic order, but with pressure-dependent Q (*23*). Further evidence for antiferromagnetism has been claimed in Ref. (*11*), in which it was claimed that $T_{m2}$ is suppressed with applied magnetic field for all field directions. It was also argued that that field-temperature phase diagram under pressure is similar to other heavy fermion antiferromagnets.

Though scattering measurements are needed to confirm the character of the high-pressure phase, the likely proximity to antiferromagnetism under pressure invites further consideration of the ambient pressure magnetic fluctuations in $UTe_2$. We note that magnetization and muon spin resonance measurements at ambient pressure exhibit scaling consistent with ferromagnetic fluctuations (*2, 24*). Nonetheless, a plot of the magnetic susceptibility times temperature (χT)

*versus* temperature indicates the dominance of antiferromagnetic correlations at low temperatures (Fig. S1).

A recent spectroscopic work on the UM$_2$Si$_2$ (M = Pd, Ni, Ru, Fe) family provides a framework for considering the effect of the valence shift towards 4+ in antiferromagnetic uranium-based materials. There, it was argued that the effect of a higher U$^{4+}$ character is to cause an overall decrease in the exchange interaction between *f* and conduction electrons (*17*). As a result, the large-moment antiferromagnetic member UPd$_2$Si$_2$ exhibits a higher 5f$^2$ contribution compared to Pauli paramagnet UFe$_2$Si$_2$. This balance is epitomized in superconducting URu$_2$Si$_2$, which becomes antiferromagnetic under applied pressure. A smaller exchange interaction under pressure promotes a smaller Kondo scale, which is expected to drive the system from a superconducting to an antiferromagnetic ground state due to the competition between Kondo and RKKY energy scales. Indeed, we see clear evidence for a suppression of the Kondo coherence temperature in UTe$_2$ as the pressure is increased (see Fig. S2). Though counterintuitive, pressure plays the opposite role as compared to CeRhIn$_5$, for which applied pressure increases the coherence temperature above 1 GPa and yields a change from an antiferromagnetic to a superconducting ground state (*25*). This is why the temperature-pressure phase diagram of UTe$_2$ mirrors that of CeRhIn$_5$.

Our results unearth a more complex interplay between superconductivity and magnetism in UTe$_2$ than previously thought, which unveil important consequences for the nature of the pairing state. Generally, spin-triplet superconductivity is expected to arise out of ferromagnetic fluctuations (*16*). The extremely large critical field (*2*), the small change in the Knight shift below T$_c$ (*4*), and the reentrant superconductivity (*5*) are indeed strong evidence for triplet pairing. Evidence for two antiferromagnetic transitions at high pressure, however, is in contradiction with a simple scenario in which ferromagnetic fluctuations solely drive the phase diagram of UTe$_2$. Interestingly, a recent theoretical study predicts that, as the on-site Coulomb interaction increases, the ground state of UTe$_2$ changes from ferromagnetic to frustrated antiferromagnetic (*26*). A particularly striking feature of the temperature-pressure phase diagram of UTe$_2$ is the fact that one of the two nearly-degenerate superconducting transition temperatures at ambient pressure is enhanced by a factor of two near a putative antiferromagnetic critical point at P$_{c1}$ = 1.3 GPa, whereas the other superconducting transition is suppressed. Such an observation would be more naturally explained if the two superconducting states at ambient pressure were a singlet and a triplet state, rather than two triplet states. Even if the magnetic order at high pressures were ferromagnetic, one would expect an enhancement of both transition temperatures if the underlying pairing states were both triplet. It would be informative to examine whether a mixed singlet-triplet state can explain the small Knight shift below T$_c$, though such state would seemingly be at odds with the observation of a trainable Kerr effect in UTe$_2$ upon application of a *c*-axis magnetic field (*7*). Applying negative chemical pressure via doping may tune the system toward a ferromagnetic ground state and provides an intriguing path for future studies to uncover the nature of the pairing state of UTe$_2$.

**Materials and Methods**
UTe$_2$ crystals were grown using a vapor transport technique with a ratio of 1U:1.5Te, as reported elsewhere (*2*). The crystallographic structure was verified by single-crystal diffraction at room temperature using Mo radiation in a commercial diffractometer, which resulted in lattice parameters a =4.1647(2) Å, b=6.1368(3) Å, c=13.9899(6) Å within the Immm (71) space group. Attempts to grow UTe$_2$ with a more stoichiometric mix of uranium and tellurium resulted either in the growth of the tetragonal phase of UTe$_2$ or a lower-quality orthorhombic phase. Zero-pressure heat capacity was measured using a Quantum Design PPMS with $^3$He option. Two single crystals of UTe$_2$ from the same growth were measured simultaneously in a piston-clamp pressure cell using Daphne oil 7373 as the pressure medium. Resistivity was measured on one crystal using a standard four-point technique with current along the [100] direction. Error bars in the inset of Fig. 2(B) are calculated from RMS noise measurements reported by the manufacturer of the resistance bridge. The zero-resistance state was determined as the point where d$\rho$/dT reached zero within experimental uncertainty. Ac calorimetry measurements were performed on the second crystal mounted in the same pressure cell (*27*). Transition temperatures from ac calorimetry were determined as the peak in C/T, except for $T_{c2}$ at 0.97 and 1.18 GPa where the transition was identified by an inflection in C/T. The pressure was calibrated using high-purity lead as a reference manometer. The lead transition remained sharp across the pressure region investigated, which indicates a hydrostatic environment. Pressure-dependent resistivity and ac specific heat were measured in a combination of $^3$He cryostat and $^4$He cryostats, as well as in an adiabatic demagnetization refrigerator.

The UTe$_2$ uranium $L_3$ XANES was measured as a function of pressure at the 4-ID-D beamline of the Advanced Photon Source, Argonne National Laboratory. High pressure was generated using a CuBe diamond anvil cell fitted with a set of 600 microns regular and partially perforated anvils, the latter used to mitigate the diamond x-ray absorption. Data were collected in transmission geometry using a pair of photodiodes to monitor the x-ray intensity before and after the sample. The x-ray energy was calibrated by measuring the yttrium *K* edge of a reference foil. While the energy was not recalibrated during the high-pressure experiments, we expect that the U $L_3$ edge shifted less than 0.3 eV during the experiment due to extrinsic factors, which is then an upper limit for any energy drift during the measurements. The pressure cell was cooled using a helium flow cryostat. The temperature was kept at 1.7(1) K during data collection, whereas it was raised to 15K during pressurization. Double stage helium gas membranes were used to control pressure *in-situ*. A stainless-steel gasket was pre-indented to about 70 microns, and a sample space of 300 microns diameter was laser drilled (*28*). Si oil was used as pressure media, and a ruby sphere as manometer. US$_2$ and UCd$_{11}$ were also measured as U$^{4+}$ and U$^{3+}$ references, respectively, although US$_2$ was later found to be an unreliable 4+ reference due to hybridization (*29*). These samples were measured in fluorescence geometry at room temperature. Data normalization was performed using the Demeter software package (*30*). The uranium absorption edge was modelled using an arctangent step function combined with a Gaussian peak (Fig. S1). This model was adjusted to the data by varying all parameters, except for the width of the step function (set to the uranium $L_3$ core-hole lifetime of 7.43 eV) (*31*), and its position (set to the maximum of the XANES first derivative).

**H2: Supplementary Materials**
Section S1. Landau theory for the relationship between the specific heat jumps of the two superconducting transitions
Fig. S1. Magnetic susceptibility times temperature (χT) versus temperature at ambient pressure.
Fig. S2. Coherence temperature.
Fig. S3. Uranium L3 absorption edge fits.

**Acknowledgments**

**General**: We would like to thank P. Orth, A. Gillman, and C.H. Booth for fruitful discussions.

**Funding:** The experimental work at Los Alamos was performed under the auspices of the U.S. Department of Energy, Office of Basic Energy Sciences, Division of Materials Science and Engineering under project "Quantum Fluctuations in Narrow-Band Systems." F. Santos was supported by FAPESP under grants No. 2016/11565-7 and 2018/20546-1. This research used resources of the Advanced Photon Source, a U.S. Department of Energy (DOE) Office of Science User Facility operated for the DOE Office of Science by Argonne National Laboratory under Contract No. DE-AC02-06CH11357. Theory work (M.H.C. and R.M.F.) was supported by the U.S. Department of Energy, Office of Science, Basic Energy Sciences, Materials Science and Engineering Division, under Award No. DE-SC0020045.

**Author contributions:** F.B. Santos, P.F.S. Rosa, and E.D. Bauer synthesized the samples. S.M. Thomas, P.F.S. Rosa, and T. Asaba prepared the samples and performed calorimetry and electrical resistivity measurements. G. Fabbris performed the XANES measurements. M.H. Christensen and R.M Fernandes developed the phenomenological model and assisted in the data interpretation. All authors contributed in the writing of the manuscript.

**Competing interests:** There are no competing interests.

**Data and materials availability:** Data are available from the corresponding author on reasonable request.

**Figures and Tables**

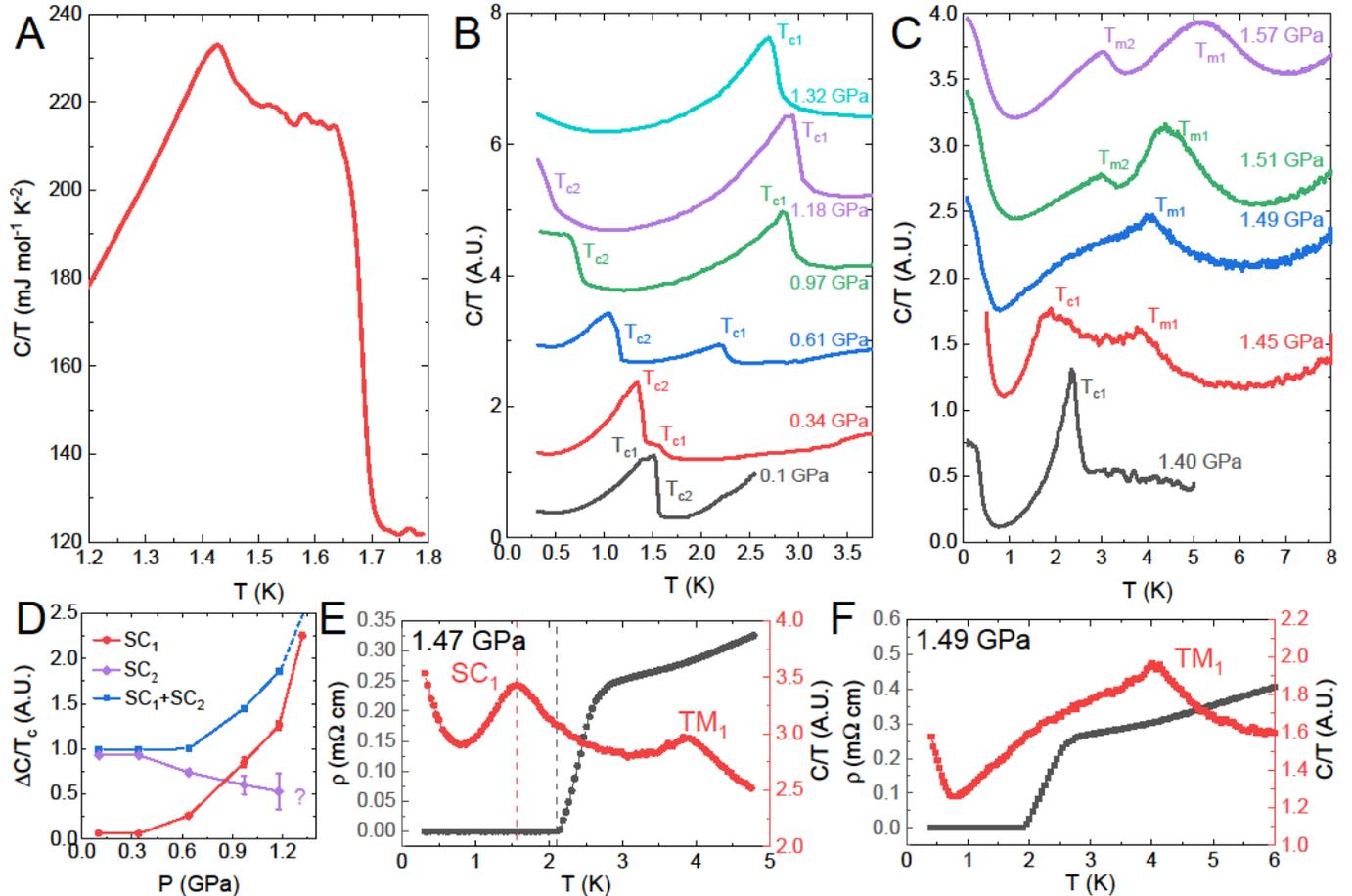

**Fig. 1. Ac calorimetry and resistivity under pressure.** (**A**) Heat capacity at zero pressure showing two superconducting transitions. (**B**) Ac calorimetry up to 1.32 GPa. $T_{c1}$ and $T_{c2}$ cross between 0.1 and 0.34 GPa. (**C**) Ac calorimetry between 1.40 GPa and 1.57 GPa. Magnetic order emerges at 1.45 GPa, splitting into two magnetic transitions at higher pressure. The low-temperature tail observed in ac calorimetry for pressures above 1.25 GPa is of unknown origin, but it is unrelated to any superconducting transition as indicated by its presence even at 1.57 GPa where the low temperature resistance does not approach zero. The curves in (B) and (C) were offset for clarity. (**D**) $\Delta C/T$ versus pressure for each of the superconducting transitions. $\psi_1$ corresponds with $T_{c1}$ and $\psi_2$ with $T_{c2}$. Specific heat jumps were determined by subtracting a baseline from each of the transitions and using the resulting peak value and temperature. (**E**) A comparison between ac calorimetry and resistivity at 1.47 GPa. The superconducting peak and the offset of the zero-resistance state differ by 0.5 K, as shown by dashed lines. (**F**) A comparison between ac calorimetry and resistivity at 1.49 GPa. A current density of 16 mA/cm² was used for the resistivity measurements in (E) and (F).

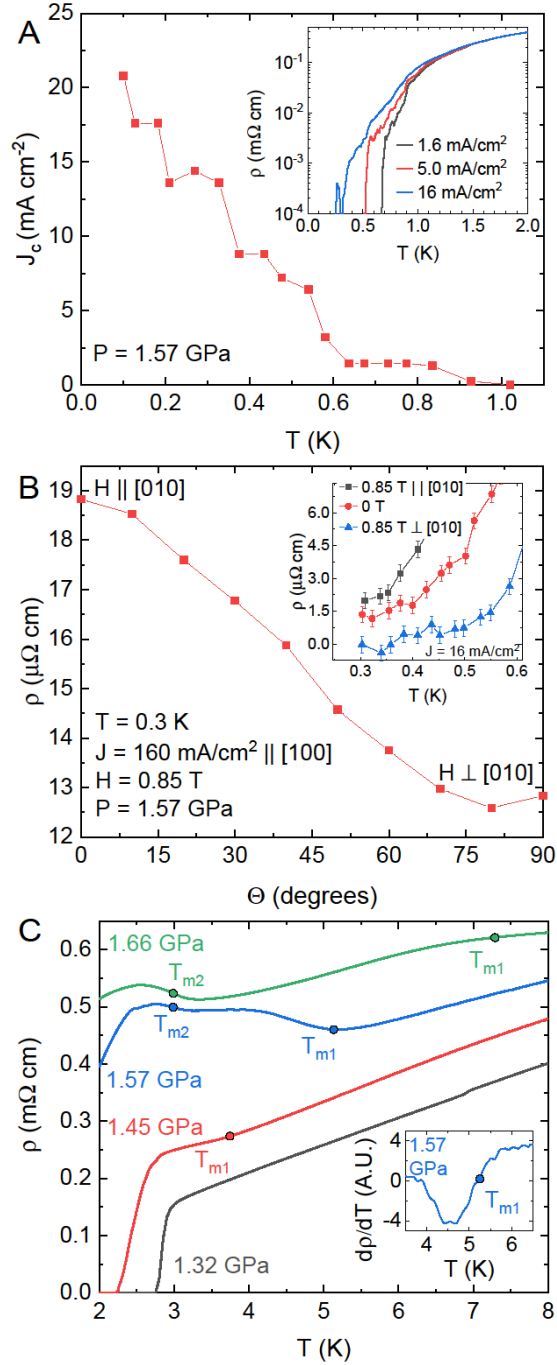

**Fig. 2. Electrical resistivity measurements.** (**A**) Critical current density versus temperature at 1.57 GPa. Inset shows resistivity versus temperature measured at different current densities. (**B**) Resistivity versus angle of applied field at 1.57 GPa as the field is rotated from parallel to [010] to perpendicular to [010]. Inset shows a comparison of resistivity versus temperature for 0 T and for 0.85 T applied either parallel or perpendicular to [010]. (**C**) Resistivity versus temperature at higher temperatures. Inset shows $d\rho/dT$ at 1.57 GPa. Current density was 160 mA/cm$^2$. Circular markers indicate transition temperatures determined from ac calorimetry.

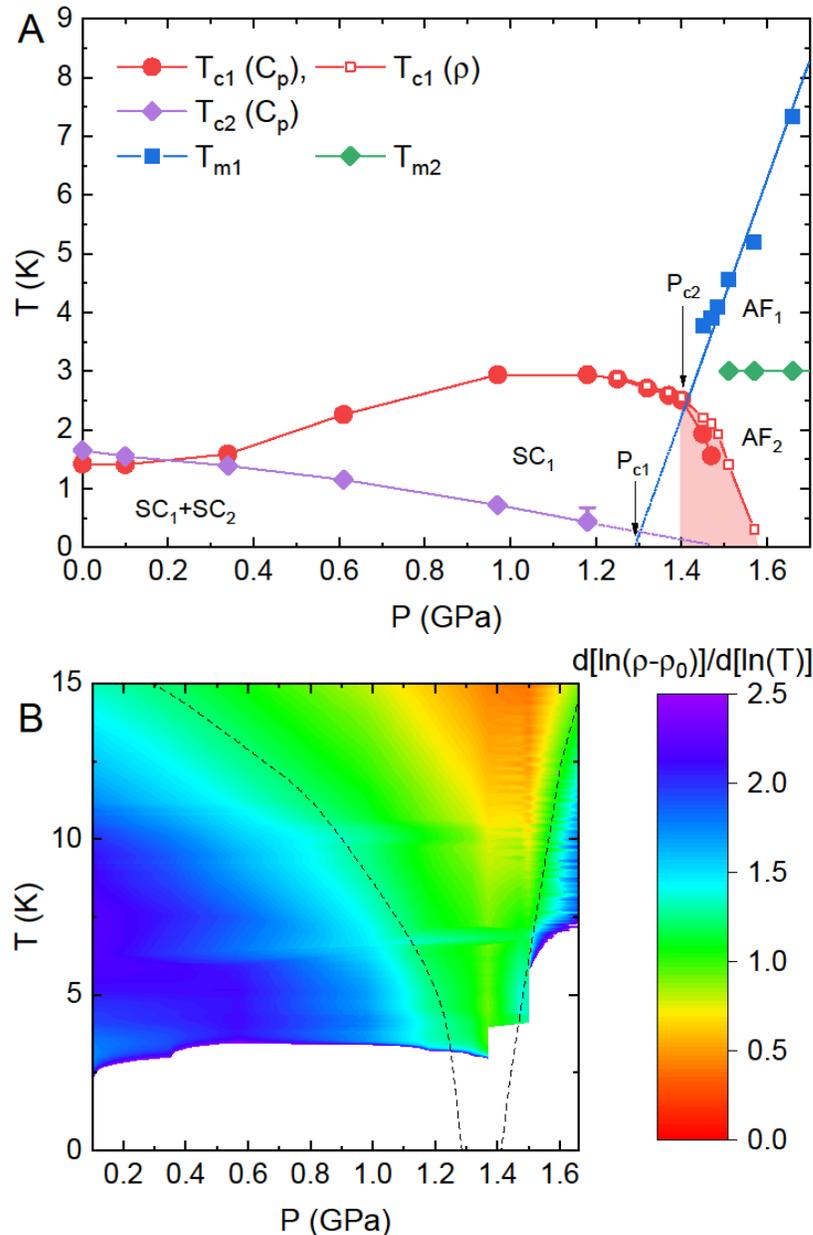

**Fig. 3. Phase diagram and electrical resistivity exponent.** (**A**) Temperature versus pressure phase diagram for UTe$_2$. Shaded area indicates the region where heat capacity and resistivity show different superconducting temperatures. Below 1.25 GPa the higher temperature superconducting transition had the same transition temperature in both heat capacity and resistivity. There is some uncertainty in the transition temperature for T$_{c2}$ at 1.18 GPa due to the potential for sample heating. T$_{c1}$ at 1.57 GPa in resistivity was determined using a current density of 16 mA/cm$^2$. (**B**) A plot of the exponent in $\rho(T) = \rho_0 + AT^n$ by taking $d[\ln(\rho - \rho_0)]/d[\ln(T)] = n$. Dotted lines are a guide to the eye for boundaries of T-linear behavior, indicating a putative quantum critical point at a pressure near 1.3 GPa.

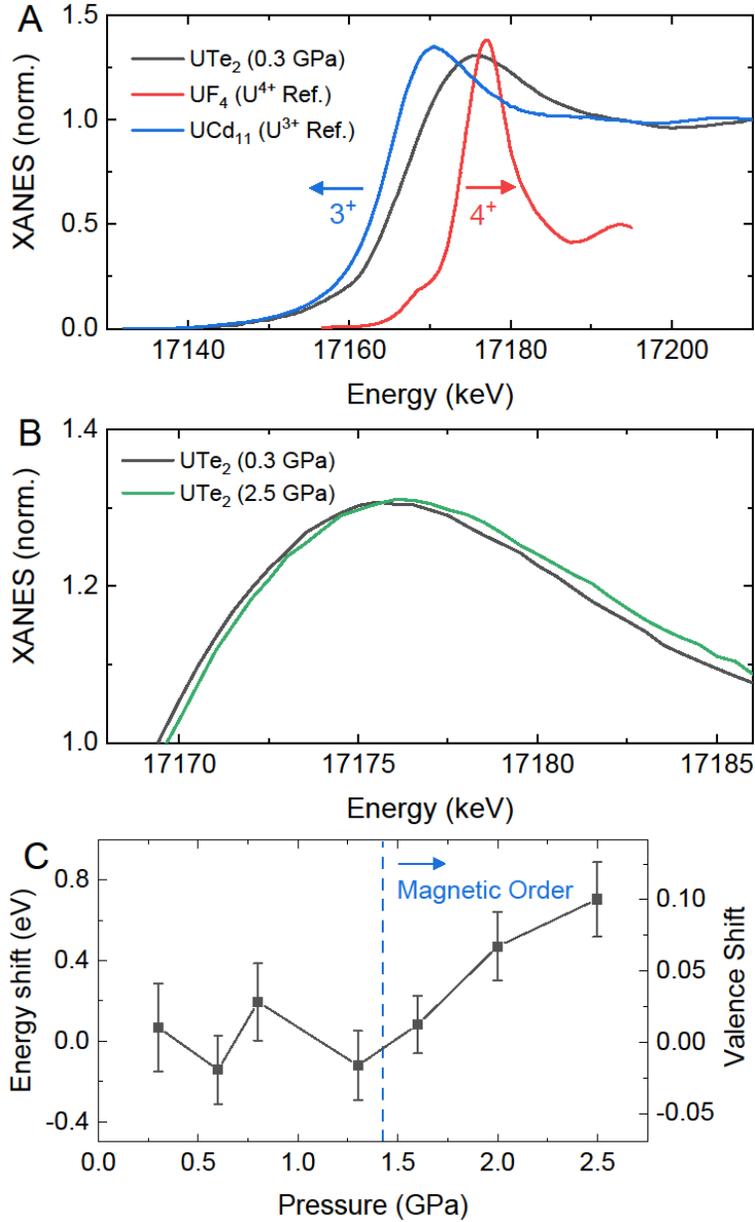

**Fig. 4. Uranium L$_3$ x-ray absorption near edge spectroscopy (XANES) spectra of UTe$_2$.** (**A**) Edge step normalized XANES data for UTe$_2$ at 0.3 GPa and reference materials UCd$_{11}$ and UF$_4$ at ambient pressure. UF$_4$ data were adapted from (*22*). (**B**) Edge step normalized XANES data for UTe$_2$ at the minimum and maximum pressures, showing a small shift towards 4$^+$ at higher pressures (**C**) The energy shift of UTe$_2$ as a function of pressure. Right axis shows estimated valence shift by taking UCd$_{11}$ and UF$_4$ as U$^{3+}$ and U$^{4+}$ references, respectively. An apparent increase in valence starts at pressures higher than 1.25 GPa.

**Supplementary Materials**

**Section S1: Landau theory for the relationship between the specific heat jumps of the two superconducting transitions**

The point group of UTe$_2$ is D$_{2h}$, which only admits one-dimensional irreducible representations (irreps). Thus, hereafter we consider two superconducting order parameters that can be fine-tuned (*e.g.*, via pressure) to have the same transition temperature T$_c$. The Landau free-energy expansion is then:

$$\mathcal{F} = \alpha(T-T_c)\psi_1^2 + \alpha\kappa(T-T_c)\psi_2^2 + \beta_1\psi_1^4 + \beta_2\psi_2^4 + 2g_1\psi_1^2\psi_2^2\cos 2\phi + g_2\psi_1^2\psi_2^2 + \epsilon(\psi_2^2 - \psi_1^2) \quad (1)$$

where $\psi_1$ and $\psi_2$ are the magnitudes of the two superconducting complex order parameters and $\phi$ is the relative phase between them. Here, $\alpha, \kappa$ are positive Landau coefficients. When the parameter $\epsilon$ is fine-tuned to zero, the two transitions take place at the same temperature T$_c$. In the phase diagram of UTe$_2$, this happens at $P^* \approx 0.2$ GPa. Therefore, we can expand $\epsilon \propto P - P^*$, such that $\epsilon < 0$ ($P < P^*$) favors the state $\psi_2$ and $\epsilon > 0$ ($P > P^*$) favors $\psi_1$. Note that this formalism does not specify the particular irreducible representations associated with $\psi_i$.

The Kerr measurements performed in Ref. (*7*) impose restrictions on the Landau parameters, since they identified, below T$_c$, a state in which both $\psi_1$ and $\psi_2$ are simultaneously non-zero and time-reversal symmetry is broken, indicative of $\phi = \pm\pi/2$. This result implies $g_1 > 0$. Thus, minimization of the free energy with respect to $\phi$ leads to:

$$\mathcal{F} = \alpha(T-T_c)\psi_1^2 + \alpha(T-T_c)\psi_2^2 + \beta_1\psi_1^4 + \beta_2\psi_2^4 - 2g\psi_1^2\psi_2^2 + \epsilon(\psi_2^2 - \psi_1^2) \quad (2)$$

with $g \equiv g_1 - \dfrac{g_2}{2}$. Moreover, for the two states to coexist microscopically, it must follow that $\beta_1\beta_2 - g^2 > 0$. Finally, we impose $\beta_1, \beta_2 > 0$ to ensure the stability of the Landau functional.

Within a mean-field approach, for $\epsilon = 0$, there is only one transition at T$_c$ to a state where both $\psi_1$ and $\psi_2$ coexist microscopically, with a relative phase of $\pm\pi/2$. The specific heat jump across this transition is:

$$\frac{\Delta C}{T_c} = \frac{\alpha^2}{2}\frac{\beta_2 + \kappa(2g + \beta_1\kappa)}{\beta_1\beta_2 - g^2} \quad (3)$$

For $\epsilon > 0$, $\psi_1$ condenses first at $T_{c,1} = T_c + \epsilon/\alpha$, followed by a secondary condensation of $\psi_2$ at:

$$T_{c,2} = T_{c,1} - \epsilon\frac{\beta_1(1+\kappa)}{\alpha(g+\beta_1\kappa)} \quad (4)$$

Note that time-reversal symmetry is only broken below $T_{c,2}$. Calculating the specific heat, we obtain two jumps at the two transitions:

$$\frac{\Delta C_1}{T_{c,1}} = \frac{\alpha^2}{2\beta_1} \frac{\Delta C_2}{T_{c,2}} \tag{5}$$

$$\frac{\Delta C_2}{T_{c,2}} = \frac{\alpha^2 (g + \beta_1 \kappa)^2}{2\beta_1 (\beta_1 \beta_2 - g^2)} \tag{6}$$

For $\epsilon < 0$, $\psi_2$ condenses first at $T_{c,2} = T_c + |\epsilon|/(\alpha\kappa)$ followed by the condensation of $\psi_1$ at $T_{c,1}$ given by:

$$T_{c,1} = T_{c,2} - |\epsilon| \frac{\beta_2(1+\kappa)}{\alpha\kappa(\beta_2 + g\kappa)} \tag{7}$$

The specific heat jumps at these two transitions are given by:

$$\frac{\Delta C_2}{T_{c,2}} = \frac{\alpha^2 \kappa^2}{2\beta_2} \tag{8}$$

$$\frac{\Delta C_1}{T_{c,1}} = \frac{\alpha^2 (\beta_2 + g\kappa)^2}{2\beta_2 (\beta_1 \beta_2 - g^2)} \tag{9}$$

Note that $\Delta C_i$ depends on $\epsilon$, since $T_{c,i}$ depends on $\epsilon$. However, $\Delta C_i / T_{c,i}$ is independent of $\epsilon$. As a result, we find the following relationship valid for any $\epsilon$:

$$\frac{\Delta C}{T_c} = \frac{\Delta C_1}{T_{c,1}} + \frac{\Delta C_2}{T_{c,2}} \tag{10}$$

Therefore, the jump divided by $T_c$ at the simultaneous transition is equal to the sum of the jumps (each divided by their respective transition temperature) when the transitions are split. In contrast, the sum of the two jumps is not the same as the jump of the simultaneous transition:

$$\Delta C_1 + \Delta C_2 = \Delta C + \epsilon \frac{\alpha(g - g\kappa + \beta_1 \kappa - \beta_2)}{2(\beta_1 \beta_2 - g^2)} \tag{11}$$

Interestingly, if $\psi_1$ and $\psi_2$ belonged to the same two-dimensional irrep, the last term would vanish since $\kappa = 1$ and $\beta_1 = \beta_2$. Although two-dimensional irreps are not supported by the D$_{2h}$ point group of UTe$_2$, this general result does provide an interesting criterion to distinguish whether coincident superconducting transitions arise from a single two-dimensional irrep or two one-dimensional irreps, which is an ongoing discussion for Sr$_2$RuO$_4$.

Although our ac calorimetry measurements do not provide quantitative values for the specific heat jumps, the results shown in Fig. 1(D) of the main text suggest that $\Delta C_1 / T_{c,1}$ and $\Delta C_2 / T_{c,2}$ (and thus their sum) depend only weakly on pressure for $P < 0.7$ GPa. This is consistent with the main assumption of our model, namely, that the main effect of pressure near the degeneracy point $P^* \approx 0.2$ GPa is to shift the transition temperatures of the two superconducting states, without significantly affecting the quartic coefficients of the Landau free-energy expansion in Eq. (1). This would imply that time-reversal symmetry breaking should take place over this pressure range.

The variation of $\Delta C_i / T_{c,i}$ with pressure, already incipient at $P = 0.6$ GPa and more pronounced for $P \geq 1$ GPa, is an indication that the impact of pressure on the quartic coefficients of Eq. (1) becomes more significant in this pressure range. Indeed, from Eq. (5), it is clear that a coefficient $\beta_1$ that increases with $\epsilon$ can cause a suppression of $\Delta C_1 / T_{c,1}$ but an enhancement of $\Delta C_2 / T_{c,2}$ depending on the values of the other Landau coefficients (the same argument would hold for Eq. (8) with $\beta_1$ replaced by $\beta_2$). Of course, there are several allowed couplings that are quartic in $\psi_i$ but linear in $\epsilon$, e.g. $\epsilon \psi_1^2 \psi_2^2$, $\epsilon(\psi_1^4 + \psi_2^4)$, and $\epsilon(\psi_1^4 - \psi_2^4)$. While determining which of these terms is the most relevant is not possible with the current data, the dependence of $\Delta C_i / T_{c,i}$ with pressure indicates that the type of coexistence between the two superconducting states may be different at high pressures as compared to ambient pressure.

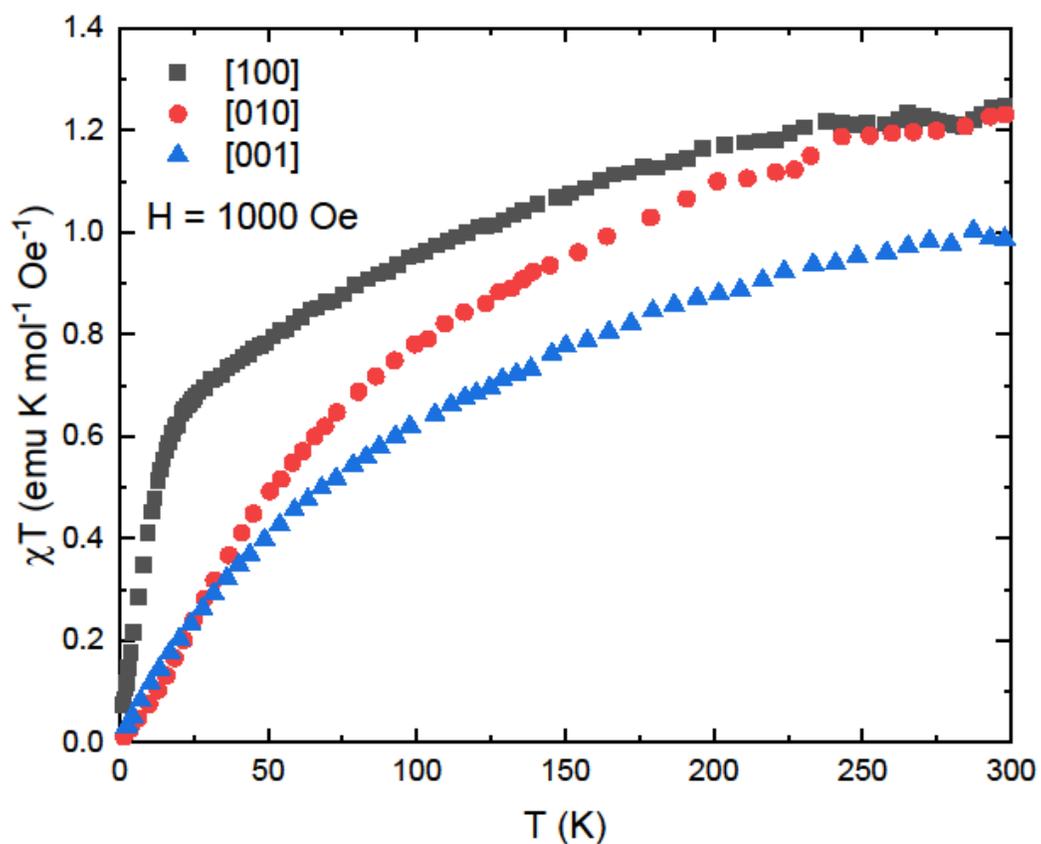

**Figure S1. Magnetic susceptibility times temperature (χT) versus temperature at ambient pressure.** Adapted from Ref. (*2*). The downward curvature at low temperatures is characteristic of dominant antiferromagnetic correlations even at ambient pressure.

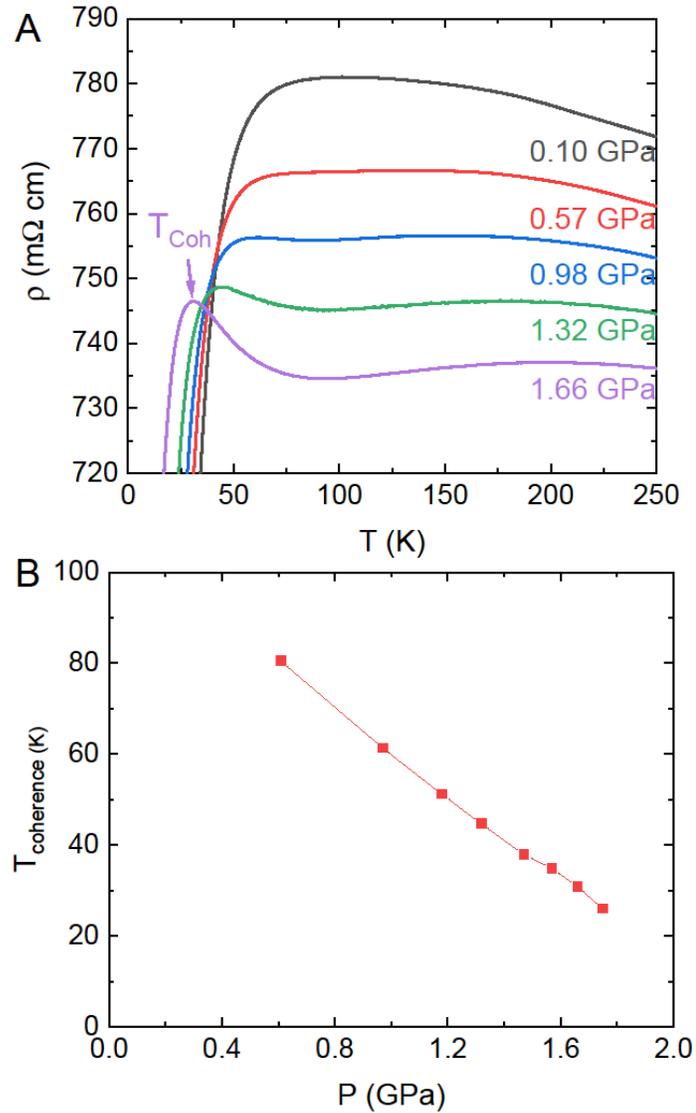

**Figure S2. Coherence temperature.** (**A**) Resistivity versus temperature as a function of pressure up to 250 K. As pressure is increased a clear peak is observed in resistivity versus temperature that indicates the temperature for Kondo cohnerece. (**B**) The Kondo coherence temperature versus pressure. As pressure is increased, the Kondo coherence is supressed to lower temperature.

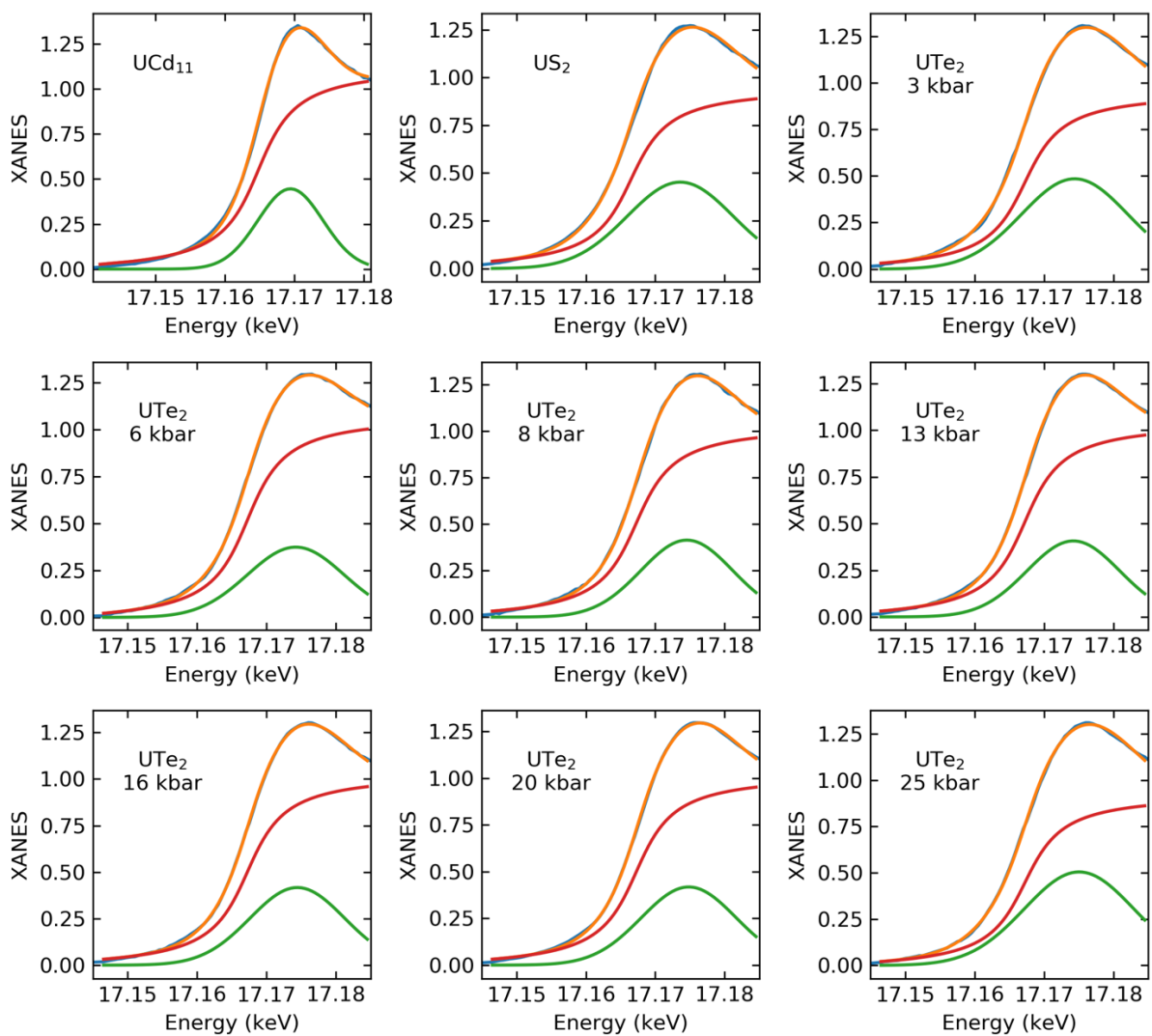

**Figure S3. Uranium L3 absorption edge fits.** The uranium L3 XANES was modelled using an arctan step function combined with a gaussian peak. The orange line is the combined fit function and overlaps with the data (blue). The red line represents the component of the fit from the arctangent step function, and the green line is the component from the Gaussian peak. Initially, $US_2$ was believed to be a reasonable $U^{4+}$ reference. Our broadened XANES spectrum, however, suggests that $US_2$ might be hybridized (*29*) and does not serve as a good $U^{4+}$ reference in comparison to $UF_4$.